\documentclass{article}
\usepackage{spconf,amsmath}

\usepackage[pdftex]{graphicx}

\usepackage{bm}
\usepackage{hyperref}
\usepackage{cleveref}

\usepackage{subcaption}

\usepackage{amsfonts}
\usepackage{xcolor}

\usepackage{upgreek}
\def\rvpi{{\bm{\uppi}}}

\newcommand{\keyword}[1]{\emph{#1}}

\usepackage{mathtools}

\usepackage{cite}

\usepackage{tabu}
\usepackage{booktabs}
\usepackage{multirow}

\def\eqref#1{equation~\ref{#1}}

\def\1{\bm{1}}

\def\rt{{\textnormal{t}}}

\def\rvy{{\mathbf{y}}}

\def\rmX{{\mathbf{X}}}

\def\vb{{\bm{b}}}

\def\vy{{\bm{y}}}

\def\mH{{\bm{H}}}

\def\mK{{\bm{K}}}

\def\mQ{{\bm{Q}}}

\def\mV{{\bm{V}}}
\def\mW{{\bm{W}}}
\def\mX{{\bm{X}}}

\DeclareMathAlphabet{\mathsfit}{\encodingdefault}{\sfdefault}{m}{sl}
\SetMathAlphabet{\mathsfit}{bold}{\encodingdefault}{\sfdefault}{bx}{n}

\def\sL{{\mathbb{L}}}

\newcommand{\R}{\mathbb{R}}

\def\vpi{{\bm{\pi}}}
\usepackage[cbgreek]{textgreek}
\def\rpi{{\textnormal{\textpi}}}

\def\lossctc{\mathcal{L}_{\text{CTC}}}
\def\HA{\textbf{HdAtt}}
\def\MHA{\textbf{MltHdAtt}}
\def\FFN{\textbf{FFN}}
\def\ReLU{\text{ReLU}}

\usepackage[hang,flushmargin]{footmisc}
\newcommand\blfootnote[1]{%
  \begingroup
  \renewcommand\thefootnote{}\footnote{#1}%
  \addtocounter{footnote}{-1}%
  \endgroup
}

\title{Self-attention networks for connectionist temporal\\
classification in speech recognition}
\name{Julian Salazar$^{\star}$ \qquad Katrin Kirchhoff$^{\star \dagger}$ \qquad Zhiheng Huang$^{\star}$}
\address{$^{\star}$Amazon AI \qquad $^{\dagger}$University of Washington \\
\normalsize \texttt{\{julsal,katrinki,zhiheng\}@amazon.com}}

\begin{document}
\ninept

\maketitle

\begin{abstract}
The success of self-attention in NLP has led to recent applications in end-to-end encoder-decoder architectures for speech recognition. Separately, connectionist temporal classification (CTC) has matured as an alignment-free, non-autoregressive approach to sequence transduction, either by itself or in various multitask and decoding frameworks. We propose \keyword{SAN-CTC}, a deep, fully self-attentional network for CTC, and show it is tractable and competitive for end-to-end speech recognition. SAN-CTC trains quickly and outperforms existing CTC models and most encoder-decoder models, with character error rates (CERs) of 4.7\% in 1 day on WSJ \keyword{eval92} and 2.8\% in 1 week on LibriSpeech \keyword{test-clean}, with a fixed architecture and one GPU. Similar improvements hold for WERs after LM decoding. We motivate the architecture for speech, evaluate position and downsampling approaches, and explore how label alphabets (character, phoneme, subword) affect attention heads and performance.
\end{abstract}


\blfootnote{\scriptsize Copyright 2019 IEEE. Published in the IEEE 2019 International Conference on Acoustics, Speech, and Signal Processing (ICASSP 2019), scheduled for 12-17 May, 2019, in Brighton, United Kingdom.
\vspace{1mm}\\
Personal use of this material is permitted. However, permission to reprint/republish this material for advertising or promotional purposes or for creating new collective works for resale or redistribution to servers or lists, or to reuse any copyrighted component of this work in other works, must be obtained from the IEEE.
\vspace{1mm}\\
\emph{Contact:} Manager, Copyrights and Permissions / IEEE Service Center / 445 Hoes Lane / P.O. Box 1331 / Piscataway, NJ 08855-1331, USA. Telephone: + Intl. 908-562-3966.}

\section{Introduction}
\label{sec:intro}

\keyword{Connectionist temporal classification} (CTC) \cite{Graves06CTC} has matured as a scalable, end-to-end approach to \emph{monotonic} sequence transduction tasks like handwriting recognition \cite{Graves09Handwriting}, action labeling \cite{Huang16ActionLabeling}, and automatic speech recognition (ASR) \cite{Graves06CTC, Graves14BLSTM, Hannun14BRDNN, Maas15BRDNN, Miao15EESEN, Amodei16DS2, Collobert16Wav2Letter, Zhang17CNNCTC, Wang17ResCTC}, sidestepping the label alignment procedure required by traditional hidden Markov model plus neural network (HMM-NN) approaches \cite{Robinson94HMMRNN}. However, the most successful end-to-end approach to \emph{general} sequence transduction has been the encoder-decoder \cite{Sutskever14S2S} with attention \cite{Bahdanau14S2SAtt}. Though first used in machine translation, its generality makes it useful to ASR as well \cite{Chan16LAS, Zhang17CNNS2S, Kim17CTCAtt, Chiu18S2SSOTA, Dong18SpeechTrans, Zeyer18S2SSOTA}. However, the lack of enforced monotonicity makes encoder-decoder ASR models difficult to train, often necessitating thousands of hours of data \cite{Chiu18S2SSOTA}, careful learning rate schedules \cite{Dong18SpeechTrans}, pretraining \cite{Zeyer18S2SSOTA}, or auxiliary CTC losses \cite{Kim17CTCAtt, Hori17CTCAttAdvances, Zeyer18S2SSOTA} to approach state-of-the-art results. The decoders are also typically autoregressive at prediction time \cite{Sutskever14S2S, Vaswani17Transformer}, restricting inference speed.

Both approaches have conventionally used recurrent layers to model temporal dependencies. As this hinders parallelization, later works proposed partially- or purely-convolutional CTC models \cite{Amodei16DS2, Collobert16Wav2Letter, Zhang17CNNCTC, Wang17ResCTC} and convolution-heavy encoder-decoder models \cite{Zhang17CNNS2S} for ASR. However, convolutional models must be significantly deeper to retrieve the same temporal receptive field \cite{Liptchinsky17Wav2Letter}. Recently, the mechanism of \keyword{self-attention} \cite{Cheng16IntraAtt, Vaswani17Transformer} was proposed, which uses the whole sequence at once to model feature interactions that are arbitrarily distant in time. Its use in both encoder-decoder and feedforward contexts has led to faster training and state-of-the-art results in translation (via the Transformer \cite{Vaswani17Transformer}), sentiment analysis \cite{Shen18DiSAN}, and other tasks. These successes have motivated preliminary work in self-attention for ASR. Time-restricted self-attention was used as a drop-in replacement for individual layers in the state-of-the-art lattice-free MMI model \cite{Povey18}, an HMM-NN system. Hybrid self-attention/LSTM encoders were studied in the context of listen-attend-spell (LAS) \cite{Sperber18SelfAttAM}, and the Transformer was directly adapted to speech in \cite{Dong18SpeechTrans, Zhou18Syllable, Zhou18Multi}; both are encoder-decoder systems.

In this work, we propose and evaluate fully self-attentional networks for CTC (\keyword{SAN-CTC}). We are motivated by practicality: self-attention could be used as a drop-in replacement in existing CTC-like systems, where only attention has been evaluated in the past \cite{Prabhavalkar17RNNTransAtt, Das18CTCAtt}; unlike encoder-decoder systems, SAN-CTC is able to predict tokens in parallel at inference time; an analysis of SAN-CTC is useful for future state-of-the-art ASR systems, which may equip self-attentive encoders with auxiliary CTC losses \cite{Kim17CTCAtt, Zeyer18S2SSOTA}. Unlike past works, we do not require convolutional frontends \cite{Dong18SpeechTrans} or interleaved recurrences \cite{Sperber18SelfAttAM} to train self-attention for ASR. In \Cref{s:math}, we motivate the model and relevant design choices (position, downsampling) for ASR. In \Cref{s:exps}, we validate SAN-CTC on the Wall Street Journal and LibriSpeech datasets by outperforming existing CTC models and most encoder-decoder models in character error rates (CERs), with fewer parameters or less training time. Finally, we train our models with different label alphabets (character, phoneme, subword), use WFST decoding to give word error rates (WERs), and examine the learned attention heads for insights.
%

\section{Model architectures for CTC and ASR}
\label{s:math}

Consider an input sequence of $T$ feature vectors, viewed as a matrix $\mX \in \R^{T \times d_{\text{fr}}}$. Let $\sL$ denote the (finite) label alphabet, and denote the output sequence as $\vy = (y_1, \dotsc, y_U) \in \sL^U$. In ASR, $\mX$ is the sequence of acoustic frames, $\sL$ is the set of graphemes/phonemes/wordpieces, and $\vy$ is the corresponding ground-truth transcription over $\sL$.

For CTC, one assumes $U \le T$ and defines an intermediate alphabet $\sL' = \sL \cup \{-\}$, where `$-$' is called the \keyword{blank symbol}. A \keyword{path} is a $T$-length sequence of intermediate labels $\vpi = (\pi_1, \dotsc, \pi_T) \in \sL'^T$. Paths are related to output sequences by a many-to-one mapping that collapses repeated labels then removes blank symbols:
\begin{equation}
\label{eq:bmap}
	\mathcal{B} : \sL'^T \to \sL^{\le T}, \ \text{e.g., } (a,b, -, -, b, b, -, a) \mapsto (a,b,b,a).
\end{equation}
In this way, paths are analogous to framewise alignments in the HMM-NN framework. CTC models the distribution of sequences by marginalizing over all paths corresponding to an output:
\begin{equation}
\label{eq:paths}
\textstyle
	P(\rvy \mid \rmX) = \sum_{\vpi \in \mathcal{B}^{-1}(\rvy)} P(\vpi \mid \rmX).
\end{equation}
Finally, CTC models each $P(\rvpi \mid \rmX)$ non-autoregressively, as a sequence of conditionally-independent outputs:
\begin{equation}
\label{eq:condind}
\textstyle
	P(\rvpi \mid \rmX) = \prod_{t=1}^T P(\rpi_t, t \mid \rmX).
\end{equation}
This model assumption means each $P(\rpi_t, t \mid \rmX)$ could be computed in parallel, after which one can do prediction via beam search, or training with gradient descent using the objective $\lossctc(\mX, \vy) = \textstyle - \log P(\vy \mid \mX)$; the order-monotonicity of $\mathcal{B}$ ensures $\lossctc$ can be efficiently evaluated with dynamic programming \cite{Graves06CTC, Graves14BLSTM}.

\subsection{Recurrent and convolutional models}

In practice, one models $P(\rvpi, \rt \mid \rmX)$ with a neural network. As inspired by HMMs, the model simplification of conditional independence can be tempered by multiple layers of (recurrent) bidirectional long short-term memory units (BLSTMs) \cite{Graves06CTC, Graves09Handwriting, Graves14BLSTM, Huang16ActionLabeling}. However, these are computationally expensive (\Cref{table:complex}), leading to simplifications like gated recurrent units (GRUs) \cite{Amodei16DS2, Zhou17PolicyCTC}; furthermore, the success of the $\ReLU(x) = \max(0, x)$ nonlinearity in preventing vanishing gradients enabled the use of vanilla bidirectional recurrent deep neural networks (BRDNNs) \cite{Hannun14BRDNN, Hannun14DS1, Maas15BRDNN} to further reduce operations per layer.

Convolutions over time and/or frequency were first used as initial layers to recurrent neural models, beginning with HMM-NNs \cite{Sainath15CLDNN} and later with CTC, where they are viewed as promoting invariance to temporal and spectral translation in ASR \cite{Amodei16DS2}, or image translation in handwriting recognition \cite{Xie16Handwriting}; they also serve as a form of dimensionality reduction (\Cref{ss:downsampling}). However, these networks were still bottlenecked by the sequentiality of operations at the recurrent layers, leading \cite{Amodei16DS2} to propose row convolutions for unidirectional RNNs, which had finite lookaheads to enable online processing while having some future context.

This led to convolution-only CTC models for long-range temporal dependencies \cite{Collobert16Wav2Letter, Zhang17CNNCTC, Wang17ResCTC}. However, these models have to be very deep (e.g., 17-19 convolutional layers on LibriSpeech \cite{Liptchinsky17Wav2Letter}) to cover the same context (\Cref{table:complex}). While in theory, a relatively local context could suffices for ASR, this is complicated by alphabets $\sL$ which violate the conditional independence assumption of CTC (e.g., English characters \cite{Zenkel18SubwordCTC}). Wide contexts also enable incorporation of noise/speaker contexts, as \cite{Sperber18SelfAttAM} suggest regarding the broad-context attention heads in the first layer of their self-attentional LAS model.

\subsection{Motivating the self-attention layer}
\label{ss:sal}

\begin{figure}[htb]
\begin{minipage}[b]{.48\linewidth}
	\centering
	\centerline{\includegraphics[height=4cm]{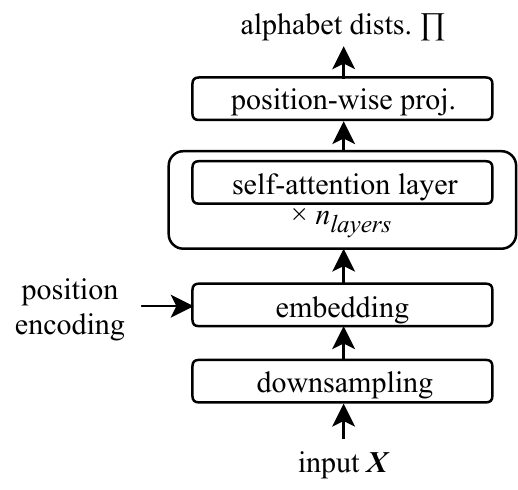}}
	\centerline{(a) SAN-CTC framework}
\end{minipage}
\begin{minipage}[b]{.48\linewidth}
	\centering
	\centerline{\includegraphics[height=4cm]{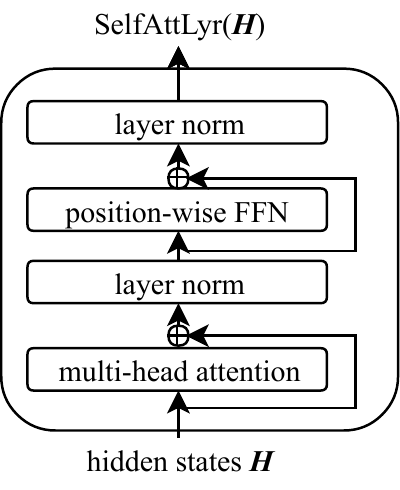}}
	\centerline{(b) Self-attention layer}
\end{minipage}
\caption{Self-attention and CTC}
\label{fig:san}
\end{figure}
\noindent We now replace recurrent and convolutional layers for CTC with self-attention \cite{Cheng16IntraAtt}. Our proposed framework (\Cref{fig:san}a) is built around \keyword{self-attention layers}, as used in the Transformer encoder \cite{Vaswani17Transformer}, previous explorations of self-attention in ASR \cite{Sperber18SelfAttAM, Dong18SpeechTrans}, and defined in \Cref{ss:salform}. The other stages are \emph{downsampling}, which reduces input length $T$ via methods like those in \Cref{ss:downsampling}; \emph{embedding}, which learns a $d_\text{h}$-dim.\ embedding that also describes token position (\Cref{ss:position}); and \keyword{projection}, where each final representation is mapped framewise to logits over the intermediate alphabet $\sL'$.

A layer's structure (\Cref{fig:san}b) is composed of two \keyword{sublayers}. The first implements \keyword{self-attention}, where the success of attention in CTC and encoder-decoder models \cite{Bahdanau14S2SAtt, Das18CTCAtt} is parallelized by using each position's representation to attend to all others, giving a contextualized representation for that position. Hence, the full receptive field is immediately available at the cost of $O(T^2)$ inner products (\Cref{table:complex}), enabling richer representations in fewer layers.

\begin{table}[htb]
\begin{minipage}[b]{1.0\linewidth}
\centering
\footnotesize
\begin{tabu}{XXXX}
	\toprule
	\textbf{Model} & \textbf{Operations per layer} & \textbf{Sequential\ operations}\ & \textbf{Maximum path length} \\
	\midrule
	Recurrent & $O(T d^2)$ & $O(T)$ & $O(T)$ \\
	Convolutional & $O(k T d^2)$ & $O(1)$ & $O(T/k)$ \\
	Convolutional (strided/dilated/pooled) & $O(k T d^2)$ & $O(1)$ & $O(\log_k(T))$ \\
	Self-attention & $O(T^2 d)$ & $O(1)$ & $O(T)$ \\
	Self-attention (restricted) & $O(k T d)$ & $O(1)$ & $O(T/k)$ \\
	\bottomrule
\end{tabu}
\end{minipage}
\caption{Operation complexity of each layer type, based on \cite{Vaswani17Transformer}. $T$ is input length, $d$ is no.\ of hidden units, and $k$ is filter/context width.}
\label{table:complex}
\end{table}

We also see inspiration from convolutional blocks: residual connections, layer normalization, and tied dense layers with ReLU for representation learning. In particular, \keyword{multi-head} attention is akin to having a number of infinitely-wide filters whose weights adapt to the content (allowing fewer ``filters'' to suffice). One can also assign interpretations; for example, \cite{Sperber18SelfAttAM} argue their LAS self-attention heads are differentiated phoneme detectors. Further inductive biases like filter widths and causality could be expressed through \keyword{time-restricted self-attention} \cite{Povey18} and \keyword{directed self-attention} \cite{Shen18DiSAN}, respectively.

\subsection{Formulation}
\label{ss:salform}

Let $\mH \in \R^{T \times d_{\text{h}}}$ denote a sublayer's input. The first sublayer performs multi-head, \keyword{scaled dot-product}, self-attention \cite{Vaswani17Transformer}. For each head $i$ of $n_{\text{hds}}$, we learn linear maps $\mW_Q^{(i)}, \mW_K^{(i)} \in \R^{d_{\text{h}} \times d_{\text{k}}}$, $\mW_V^{(i)} \in \R^{d_{\text{h}} \times d_{\text{h}} / n_{\text{hds}}}$. Left multiplication by $\mH$ give the \keyword{queries} $\mQ^{(i)}$, \keyword{keys} $\mK^{(i)}$, and \keyword{values} $\mV^{(i)}$ of the $i$-th head, which combine to give
\begin{align}
	\HA^{(i)} = \sigma \left(\mQ^{(i)}  \mK^{(i)\top} / \sqrt{d_{\text{h}}}\right) \mV^{(i)},
\end{align}
where $\sigma$ is row-wise softmax. Heads are concatenated along the $d_{\text{h}} / n_{\text{hds}}$ axis to give $\MHA = [\HA^{(1)}, \dotsc, \HA^{(n_{\text{hds}})}]$. The second sublayer is a \keyword{position-wise feed-forward network} \cite{Vaswani17Transformer} $\textbf{FFN}(\mH) = \ReLU(\mH \mW_1 + \vb_1)\mW_2 + \vb_2$ where parameters $\mW_1 \in \R^{d_{\text{h}} \times d_{\text{ff}}}$, $\vb_1 \in \R^{d_{\text{ff}}}$, $\mW_2 \in \R^{d_{\text{ff}} \times d_{\text{h}}}$, $\vb_2 \in \R^{d_{\text{h}}}$ are learned, with the biases $\vb_1, \vb_2$ broadcasted over all $T$ positions. This sublayer aggregates the multiple heads at time $t$ into the attention layer's final output at $t$. All together, the layer is given by:
\begin{align}
\textbf{MidLyr}(\mH) &= \text{LN}(\MHA(\mH) + \mH),\\
\textbf{SelfAttLyr}(\mH) &= \text{LN}(\FFN(\textbf{MidLyr}(\mH)) + \textbf{MidLyr}(\mH)).
\end{align}

\subsection{Downsampling}
\label{ss:downsampling}

In speech, the input length $T$ of frames can be many times larger than the output length $U$, in contrast to the roughly word-to-word setting of machine translation. This is especially prohibitive for self-attention in terms of memory: recall that an attention matrix of dimension $\mQ^{(i)} \mK^{(i)\top} \in \R^{T \times T}$ is created, giving the $T^2$ factor in \Cref{table:complex}. A convolutional frontend is a typical downsampling strategy \cite{Amodei16DS2, Dong18SpeechTrans}; however, we leave integrating other layer types into SAN-CTC as future work. Instead, we consider three fixed approaches, from least- to most-preserving of the input data: \keyword{subsampling}, which only takes every $k$-th frame; \keyword{pooling}, which aggregates every $k$ consecutive frames via a statistic (average, maximum); \keyword{reshaping}, where one concatenates $k$ consecutive frames into one \cite{Sperber18SelfAttAM}. Note that CTC will still require $U \le T/k$, however.

\subsection{Position}
\label{ss:position}

Self-attention is inherently content-based \cite{Vaswani17Transformer}, and so one often encodes position into the post-embedding vectors. We use standard trigonometric embeddings, where for $0 \le i \le d_{\text{emb}}/2$, we define
\begin{align*}
\textstyle
	\text{PE}(t, 2i) = \sin \frac{t}{10000^{2i / d_\text{emb}}}, \ \text{PE}(t, 2i+1) = \cos \frac{t}{10000^{2i / d_\text{emb}}}.
\end{align*}
for position $t$. We consider three approaches: \keyword{content-only} \cite{Hori17CTCAttAdvances}, which forgoes position encodings; \keyword{additive} \cite{Dong18SpeechTrans}, which takes $d_{\text{emb}} = d_{\text{h}}$ and adds the encoding to the embedding; and \keyword{concatenative}, where one takes $d_{\text{emb}} = 40$ and concatenates it to the embedding. The latter was found necessary for self-attentional LAS \cite{Sperber18SelfAttAM}, as additive encodings did not give convergence. However, the monotonicity of CTC is a further positional inductive bias, which may enable the success of content-only and additive encodings.

\begin{figure*}[tb]
\centering
\begin{subfigure}[b]{0.24\textwidth}
  \includegraphics[width=1\columnwidth]{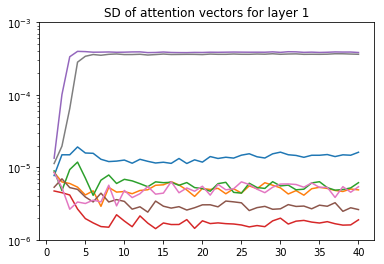}
\end{subfigure}~
\begin{subfigure}[b]{0.24\textwidth}
\centering
  \includegraphics[width=1\columnwidth]{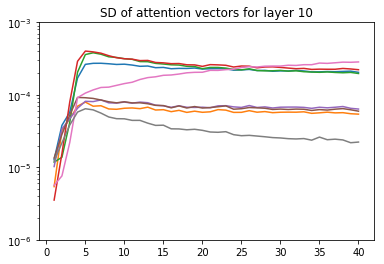}
\end{subfigure}~
\begin{subfigure}[b]{0.24\textwidth}
\centering
  \includegraphics[width=1\columnwidth]{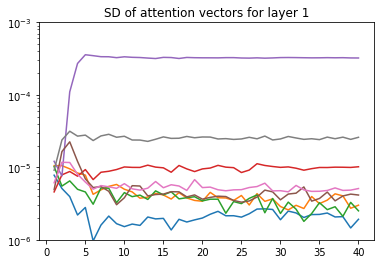}
\end{subfigure}~
\begin{subfigure}[b]{0.24\textwidth}
\centering
  \includegraphics[width=1\columnwidth]{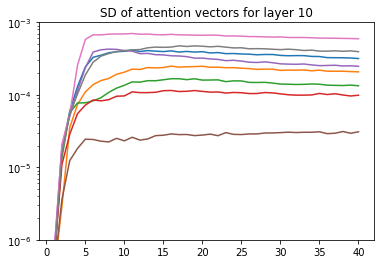}
\end{subfigure}

\begin{subfigure}[b]{0.24\textwidth}
\centering
  \includegraphics[width=1\columnwidth]{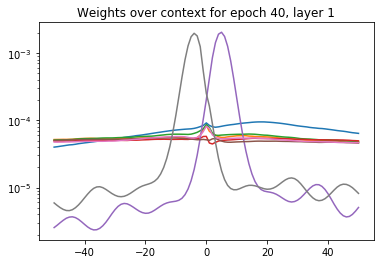}
\end{subfigure}~
\begin{subfigure}[b]{0.24\textwidth}
\centering
  \includegraphics[width=1\columnwidth]{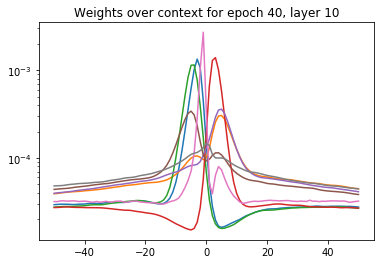}
\end{subfigure}~
\begin{subfigure}[b]{0.24\textwidth}
\centering
  \includegraphics[width=1\columnwidth]{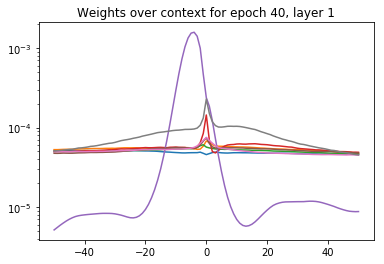}
\end{subfigure}~
\begin{subfigure}[b]{0.24\textwidth}
\centering
  \includegraphics[width=1\columnwidth]{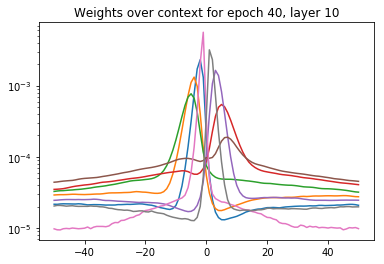}
\end{subfigure}

\caption{Average attention weights over WSJ's test set. The left half is a character model; the right half is a phoneme model. Each curve corresponds to a head, at layers 1 and 10, of representative SAN-CTC models (reshape + additive embedding). The first row charts standard deviation over time; the second row charts the attention magnitudes relative to position, where 0 represents attending to the same position.}
\label{fig:vars}
\end{figure*}
\section{Experiments}
\label{s:exps}

We take $(n_{\text{layers}},d_{\text{h}},n_{\text{heads}},d_{\text{ff}}) =$ (10, 512, 8, 2048), giving {\raise.17ex\hbox{$\scriptstyle\sim$}}30M parameters. This is on par with models on WSJ (10-30M) \cite{Graves14BLSTM, Hannun14BRDNN, Collobert16Wav2Letter} and an order of magnitude below models on LibriSpeech (100-250M) \cite{Amodei16DS2, Liptchinsky17Wav2Letter}. We use MXNet \cite{Chen15MXNet} for modeling and Kaldi/EESEN \cite{Povey11Kaldi, Miao15EESEN} for data preparation and decoding. Our self-attention code is based on GluonNLP's implementation. At train time, utterances are sorted by length: we exclude those longer than 1800 frames ($\ll$1\% of each training set). We take a window of 25ms, a hop of 10ms, and concatenate cepstral mean-variance normalized features with temporal first- and second-order differences.\footnote{Rescaling so that these differences also have var.\;$\approx1$ helped WSJ training.} We downsample by a factor of $k=3$ (this also gave an ideal $T/k \approx d_{\text{h}}$ for our data; see \Cref{table:complex}).

We perform Nesterov-accelerated gradient descent on batches of 20 utterances. As self-attention architectures can be unstable in early training, we clip gradients to a global norm of 1 and use the standard linear warmup period before inverse square decay associated with these architectures \cite{Vaswani17Transformer, Dong18SpeechTrans}. Let $n$ denote the global step number of the batch (across epochs); the learning rate is given by
	\begin{equation}
		\textstyle
		\text{LR}(n) = \frac{\lambda}{\sqrt{d_\text{h}}} \min\left(\frac{n}{n_{\text{warmup}}^{1.5}}, \frac{1}{\sqrt{n}}\right),
	\end{equation}
	 where we take $\lambda =$ 400 and $n_{\text{warmup}}$ as a hyperparameter. However, such a decay led to early stagnation in validation accuracy, so we later divide the learning rate by $10$ and run at the decayed rate for 20 epochs. We do this twice, then take the epoch with the best validation score. Xavier initialization gave validation accuracies of zero for the first few epochs, suggesting room for improvement. Like previous works on self-attention, we apply label smoothing (see Tables~\ref{table:wsj-tok-results},~\ref{table:wsj-phn-results},~\ref{table:libri-tok-results}; we also tried model averaging to no gain). To compute word error rates (WERs), we use the dataset's provided language model (LM) as incorporated by WFST decoding \cite{Miao15EESEN} to bridge the gap between CTC and encoder-decoder frameworks, allowing comparison with known benchmarks and informing systems that incorporate expert knowledge in this way (e.g., via a pronunciation lexicon).

\subsection{Wall Street Journal (WSJ)}

\begin{table}
\begin{minipage}{1.0\linewidth}
	\centering
	\footnotesize
    \begin{tabu}{Xcccc}
    \toprule
    \multirow{2}{*}{\textbf{Model}} & \multicolumn{2}{c}{\textbf{dev93}} & \multicolumn{2}{c}{\textbf{eval92}} \\
     & CER & WER & CER & WER \\
    \midrule
	CTC (BRDNN) \cite{Hannun14BRDNN} & --- & --- & 10.0 & --- \\
    CTC (BLSTM) \cite{Graves14BLSTM} & --- & --- & 9.2 & --- \\
    CTC (BLSTM) \cite{Kim17CTCAtt} & 11.5 & --- & 9.0 & --- \\
    Enc-Dec (4-1) \cite{Kim17CTCAtt} & 12.0 & --- & 8.2 & --- \\
    Enc-Dec+CTC (4-1) \cite{Kim17CTCAtt} & 11.3 & --- & 7.4 & --- \\
    Enc-Dec (4-1) \cite{Bahdanau16S2S} & --- & --- & 6.4 & 9.3 \\
    CTC/ASG (Gated CNN) \cite{Zeghidour18E2ERaw} & 6.9 & 9.5 & 4.9 & 6.6 \\
    Enc-Dec (2,1,3-1) \cite{Sabour18OCD} & --- & --- & \textbf{3.6} & --- \\
    \midrule
    CTC (SAN), reshape, additive & 7.1 & 9.3 & 5.1 & 6.1 \\
    + label smoothing, $\lambda =$ 0.1 & 6.4 & 8.9 & 4.7 & 5.9 \\
    \bottomrule
    \end{tabu}
\end{minipage}
\caption{End-to-end, MLE-based, open-vocab.\ models trained on WSJ. Only WERs incorporating the extended 3-gram LM or a 4-gram LM (Gated CNN) are listed.}
\label{table:wsj-tok-results}
\end{table}

We train both character- and phoneme-label systems on the 80-hour WSJ training set to validate our architectural choices. Similar to \cite{Kim17CTCAtt, Dong18SpeechTrans}, we use 40-dim.\ mel-scale filter banks and hence 120-dim.\ features. We warmup for 8000 steps, use a dropout of 0.2, and switch schedules at epoch 40. For the WSJ dataset, we compare with similar MLE-trained, end-to-end, open-vocabulary systems in \Cref{table:wsj-tok-results}. We get an \emph{eval92} CER of 4.7\%, outdoing all previous CTC-like results except 4.6\% with a trainable frontend \cite{Zeghidour18E2ERaw}. We use the provided extended 3-gram LM to retrieve WERs. For phoneme training, our labels come from the CMU pronunciation lexicon (\Cref{table:wsj-phn-results}).  These models train in one day (Tesla V100), comparable to the Speech Transformer \cite{Dong18SpeechTrans}; however, SAN-CTC gives further benefits at inference time as token predictions are generated in parallel.

\begin{table}
\begin{minipage}{1.0\linewidth}
	\centering
	\footnotesize
    \begin{tabu}{Xcccc}
	    \toprule
	    \multirow{2}{*}{\textbf{Model}} & \multicolumn{2}{c}{\textbf{dev93}} & \multicolumn{2}{c}{\textbf{eval92}} \\
	    & PER & WER & PER & WER \\
	    \midrule
	    CTC (BRDNN) \cite{Miao15EESEN} & --- & --- & --- & 7.87 \\
	    CTC (BLSTM) \cite{Wang17ResCTC} & --- & 9.12 & --- & 5.48 \\
	    CTC (ResCNN) \cite{Wang17ResCTC} & --- & 9.99 & --- & 5.35 \\
	    Ensemble of 3 (voting) \cite{Wang17ResCTC} & --- & \textbf{7.65} & --- & \textbf{4.29} \\
	    \midrule
CTC (SAN), reshape, additive & 7.12 & 8.09 & 5.07 & 4.84 \\
+ label smoothing, $\lambda =$ 0.1 & 6.86 & 8.16 & 4.73 & 5.23 \\
	    \bottomrule
    \end{tabu}
\end{minipage}
\caption{CTC phoneme models with WFST decoding on WSJ.}
\label{table:wsj-phn-results}
\end{table}

We also evaluate design choices in \Cref{table:wsj-ablations}. Here, we consider the effects of downsampling and position encoding on accuracy for our fixed training regime. We see that unlike self-attentional LAS \cite{Sperber18SelfAttAM}, SAN-CTC works respectably even with no position encoding; in fact, the contribution of position is relatively minor (compare with \cite{Hori17CTCAttAdvances}, where location in an encoder-decoder system improved CER by 3\% absolute). Lossy downsampling appears to preserve performance in CER while degrading WER (as information about frame transitions is lost). We believe these observations align with the monotonicity and independence assumptions of CTC.

\begin{table}
\begin{minipage}{1.0\linewidth}
	\centering
	\footnotesize
    \begin{tabu}{XXcc}
	    \toprule
	    \multirow{2}{*}{\textbf{Downsampling}} & \multirow{2}{*}{\textbf{Position embedding}} & \multicolumn{2}{c}{\textbf{dev93}} \\
	    & &  CER & WER \\
	    \midrule
	    reshape & content-only & 7.62 & 9.57 \\
	    reshape & additive & 7.10 & \textbf{9.27}\\
		reshape & concatenative & 7.10 & 9.97\\
		\midrule
	    pooling (maximum) & additive & 7.15 & 10.72 \\
	    pooling (average) & additive & \textbf{6.82} & 9.41\\
	    subsample & additive & none & none  \\
	    \bottomrule
    \end{tabu}
\end{minipage}
\caption{SAN-CTC character model on WSJ with modifications.}
\label{table:wsj-ablations}
\end{table}

Inspired by \cite{Sperber18SelfAttAM}, we plot the standard deviation of attention weights for each head as training progresses; see \Cref{fig:vars} for details. In the first layers, we similarly observe a differentiation of variances, along with wide-context heads; in later layers, unlike \cite{Sperber18SelfAttAM} we still see mild differentiation of variances. Inspired by \cite{Povey18}, we further plot the attention weights relative to the current time position (here, per head). Character labels gave forward- and backward-attending heads (incidentally, averaging these would retrieve the bimodal distribution in \cite{Povey18}) at all layers. This suggests a gradual expansion of context over depth, as is often engineered in convolutional CTC. This also suggests possibly using fewer heads, directed self-attention \cite{Shen18DiSAN}, and restricted contexts for faster training (\Cref{table:complex}). Phoneme labels gave a sharp backward-attending head and more diffuse heads. We believe this to be a symptom of English characters being more context-dependent than phonemes (for example, emitting `tt' requires looking ahead, as `--' must occur between two runs of `t' tokens).

\subsection{LibriSpeech}

We give the first large-scale demonstration of a fully self-attentional ASR model using the LibriSpeech ASR corpus \cite{Panayotov15LibriSpeech}, an English corpus produced from audio books giving 960 hours of training data. We use 13-dim.\ mel-freq.\ cepstral coeffs.\ and hence 39-dim.\ features. We double the warmup period, use a dropout of 0.1, and switch schedules at epoch 30. Using character labels, we attained a \keyword{test-clean} CER of 2.8\%, outdoing all previous end-to-end results except OCD training \cite{Sabour18OCD}. We use the provided 4-gram LM via WFST to compare WERs with state-of-the-art, end-to-end, open-vocabulary systems in \Cref{table:libri-tok-results}. At this scale, even minor label smoothing was detrimental. We run 70 epochs in slightly over a week (Tesla V100) then choose the epoch with the best validation score for testing. For comparison, the best CTC-like architecture \cite{Liptchinsky17Wav2Letter} took 4-8 weeks on 4 GPUs for its results.\footnote{\scriptsize \url{https://github.com/facebookresearch/wav2letter/issues/11}} The Enc-Dec+CTC model is comparable, taking almost a week on an older GPU (GTX 1080 Ti) to do its $\sim$12.5 full passes over the data.\footnote{\scriptsize \url{https://github.com/rwth-i6/returnn-experiments/tree/master/2018-asr-attention/librispeech/full-setup-attention}}

\begin{table}
\begin{minipage}[b]{1.0\linewidth}
	\centering
	\footnotesize
    \begin{tabu}{@{}Xccccc@{}}
	    \toprule
	    \multirow{2}{*}{\textbf{Model}} & \multirow{2}{*}{\textbf{Tok.}} & \multicolumn{2}{c}{\textbf{test-clean}} & \multicolumn{2}{c}{\textbf{test-other}} \\
	     & & CER & WER & CER & WER \\
	    \midrule
	    CTC/ASG (Wav2Letter) \cite{Collobert16Wav2Letter} & chr. & 6.9 & 7.2 & --- & --- \\
	    CTC (DS1-like) \cite{Hannun14DS1, Mozilla17DS1} & chr. & --- & 6.5 & --- & --- \\
	   	Enc-Dec (4-4) \cite{Liang18IRL} & chr. & 6.5 & --- & 18.1 & ---\\
	   	Enc-Dec (6-1) \cite{Sriram18ColdFusion} & chr. & 4.5 & --- & 11.6 & ---\\
		CTC (DS2-like) \cite{Amodei16DS2, Zhou17PolicyCTC} & chr. & --- & 5.7 & --- & 15.2 \\
	   	Enc-Dec+CTC (6-1, pretr.) \cite{Zeyer18S2SSOTA} & 10k & --- & 4.8 & --- & 15.3\\
		CTC/ASG (Gated CNN) \cite{Liptchinsky17Wav2Letter} & chr. & --- & 4.8 & --- & 14.5 \\
	    Enc-Dec (2,6-1) \cite{Sabour18OCD} & 10k & 2.9 & --- & \textbf{8.4} & ---\\
	    \midrule
	    CTC (SAN), reshape, additive & chr. & 3.2 & 5.2 & 9.9 & 13.9 \\
	    + label smoothing, $\lambda =$ 0.05 & chr. & 3.5 & 5.4 & 11.3 & 14.5 \\
	    CTC (SAN), reshape, concat. & chr. & \textbf{2.8} & 4.8 & 9.2 & 13.1 \\
	    \bottomrule
    \end{tabu}
\end{minipage}
\caption{End-to-end, MLE-based, open-vocab.\ models trained on LibriSpeech. Only WERs incorporating the 4-gram LM are listed.}
\label{table:libri-tok-results}
\end{table}

Finally, we trained the same model with BPE subwords as CTC targets, to get more context-independent units \cite{Zenkel18SubwordCTC}. We did 300 merge operations (10k was unstable) and attained a CER of 7.4\%. This gave a WER of 8.7\% with no LM (compare with \Cref{table:libri-tok-results}'s LM-based entries), and 5.2\% with a subword WFST of the LM. We still observed  attention heads in both directions in the first layer, suggesting our subwords were still more context-dependent than phonemes.

\section{Conclusion}

We introduced SAN-CTC, a novel framework which integrates a fully self-attentional network with a connectionist temporal classification loss. We addressed the challenges of adapting self-attention to CTC and to speech recognition, showing that SAN-CTC is competitive with or outperforms existing end-to-end models on WSJ and LibriSpeech. Future avenues of work include multitasking SAN-CTC with other decoders or objectives, and streamlining network structure via directed or restricted attention.

\bibliographystyle{IEEEbib}


\bibliography{salazar}

\end{document}